\documentclass{book}
\pdfoutput=1

\font\rs=cmss10.360pk
\font\rt=cmss9.360pk
\font\sd=cmcsc9.360pk

{{\small\rt \hfill Derivation of breather solutions \hfill}}

\include {mak}
\usepackage{amsmath}
\usepackage{amssymb}
\usepackage{graphicx}
\usepackage{subfigure}
\def\sech{\mathop{\rm sech}\nolimits}

\begin{document}

%Your \newcommands below (if any):

\oddsidemargin 16.5truemm
\evensidemargin 16.5truemm

\thispagestyle{plain}

\noindent{\rs Proceedings of SEAMS-GMU Conference 2007}

\vspace{-0.25cc}

\noindent{\scriptsize Applied Mathematics, pp.~357--368.}

\vspace{5cc}

\begin{center}
{\Large\bf DERIVATION OF THE NLS BREATHER SOLUTIONS USING
DISPLACED PHASE-AMPLITUDE VARIABLES
\rule{0mm}{6mm}\renewcommand{\thefootnote}{}\footnotetext{\hspace{-0.55cm}\scriptsize
\\
}\par}

\vspace{2cc}

{\large\sc Natanael Karjanto and E. van Groesen}

\vspace{2cc}

\parbox{24cc}{{\scriptsize{\bf Abstract.}
Breather solutions of the nonlinear Schr\"{o}dinger equation are
derived in this paper: the Soliton on Finite Background, the Ma
breather and the rational breather. A special Ansatz of a
displaced phase-amplitude equation with respect to a background is
used as has been proposed by van Groesen et. al. (2006).
Requiring the displaced phase to be temporally independent, has as
consequence that the dynamics at each position is described by the
motion of a nonlinear autonomous oscillator in a potential energy
that depends on the phase and on the spatial phase change. The
relation among the breather solutions is confirmed by explicit
expressions, and illustrated with the amplitude amplification
factor. Additionally, the corresponding physical wave field is
also studied and wavefront dislocation together with phase
singularity at vanishing amplitude are observed in all three
cases. \par}}
\end{center}

{\scriptsize {\it Key words and Phrases}: displaced
phase-amplitude variables, nonlinear Schr\"{o}dinger equation,
breather solutions, amplitude amplification factor, wavefront
dislocation. \par}

\vspace{1.5cc}

\begin{center}
{\bf 1. INTRODUCTION}
\end{center}

The nonlinear Schr\"{o}dinger (NLS) equation is a nonlinear
dispersive partial differential equation that has been used as a
mathematical model in many areas of Mathematical Physics, such as
nonlinear water waves, nonlinear optics and plasma physics
(Ablowitz and Segur, 1981; Ablowitz and Clarkson, 1991). In this
letter, we will consider the spatial NLS equation, given as
follows:
\begin{equation}
\partial_{\xi}\psi + i \beta \partial_{\tau}^{2}\psi + i \gamma |\psi|^{2} \psi = 0. \label{spatialNLS}
\end{equation}
We will derive and study three exact solutions of this equation.
These exact solutions are known in the literature as `breather'
solutions of the NLS equation (Dysthe and Trulsen, 1999; Dysthe,
2001; Grimshaw et. al., 2001). The name `breather' reflects the
behavior of the profile which is periodic in time or space and
localized in space or time. The concept was introduced by
Ablowitz, Kaup, Newell and Segur (Ablowitz, et. al., 1974) in the context of the sine-Gordon partial differential equation. The breather solutions
are also found in other equations, for instance in the
Davey-Stewartson equation (Tajiri and Arai, 2000) and modified
Korteweg-de Vries (mKdV) equation (Drazin and Johnson, 1989).

To find the breather solutions of the NLS equation, we use a
special Ansatz in the description with displaced phase-amplitude
variables introduced by van Groesen et. al. (2006), with a
displacement that depends on the background. Requiring the
displaced phase to be temporally independent, has as consequence
that the dynamics at each position is described by the motion of a
nonlinear autonomous oscillator in a potential energy that depends
on the phase and on the spatial phase change. It is remarkable
that this assumption leads to the breather solutions of the NLS
equation that are known in the literatures: the Soliton on Finite
Background (SFB) (Akhmediev et. al., 1987), the Ma breather (Ma,
1979) and the rational breather (Peregrine, 1983). These solutions
have been suggested as models for a class of freak wave events
seen in $2 + 1$ dimensional simulations of surface gravity waves
(Henderson et. al., 1999). However, the SFB shows many properties
of extreme, or freak, rogue wave events, as observed by Osborne
et. al. (2000), Calini and Schober (2002). The SFB is also a good
candidate in order to generate extreme wave events in a
hydrodynamic laboratory (Andonowati et. al., 2006; Huijsmans et.
al., 2005).

This letter is organized as follows. Section 2 presents some exact
solutions of the NLS equation. We demonstrate that three different
solutions of the NLS equation are obtained within a similar
formulation. Section 3 explains briefly the relationship among the
solutions of the NLS equation. Section 4 will discuss the
evolution of breather solutions in the Argand diagram. In Section
5 we will deal with surface water waves and observe the occurrence
of wavefront dislocations in the corresponding density plots of
the breather solutions. The final section draws some conclusions
and remarks on the subject in this letter.

\vspace{1.5cc}

\begin{center}
{\bf 2. EXACT SOLUTIONS OF THE NLS EQUATION}
\end{center}

The simplest nontrivial solution of the NLS equation is the
\textit{plane-wave} or the `continuous wave' (cw) solution. It
does not depend on the temporal variable $\tau$ and is given by
$A_{0}(\xi) = r_{0} e^{-i\gamma r_{0}^{2}\xi}$. Another simple
solution is the `one-soliton' or `single-soliton' solution. It can
be found by the inverse scattering technique (Zakharov and Shabat,
1972), but also much simpler by seeking a travelling-wave
solution. An explicit expression is given by
\begin{equation}
  A(\xi,\tau) = A_{0}(\xi) \sqrt{2} \sech \left(r_{0} \sqrt{\frac{\gamma}{\beta}} \tau\right).%
  \label{onesoliton}
\end{equation}
It will become clear when we consider the corresponding physical
wave field that both the cw and the one-soliton solutions have
coherent structures, while other solutions that we will deal with
below do not.

In van Groesen et. al. (2006), a displaced phase-amplitude
description has been introduced. As a special case, we will look
for a solution of the NLS equation in the form
\begin{equation}
  A(\xi,\tau) = A_{0}(\xi) [G(\xi,\tau) e^{i\phi(\xi)} - 1],
  \label{NLSsolution}
\end{equation}
for which the phase $\phi$, which is `displaced' since the
background contribution is taken out, is taken to depend only on
the spatial variable $\xi$. Substituting (\ref{NLSsolution}) into
the NLS equation (\ref{spatialNLS}), we obtain two equations that
correspond to the real and the imaginary parts, respectively.

Multiplying the real part by $\cos \phi$ and the imaginary part by
$\sin \phi$, adding both equations will give a Riccati-like
equation for $G$:
\begin{equation}
  \partial_{\xi} G + \gamma r_{0}^{2} \sin 2\phi G - \gamma
  r_{0}^{2} \sin \phi G^{2} = 0. \label{riccati}
\end{equation}
Solving this equation leads us to a conclusion that $G$ can be written in a special form. By letting $G = 1/H$, then equation \eqref{riccati} becomes a first order linear differential equation in $H$:
\begin{equation}
  \partial_{\xi} H - \gamma r_{0}^{2} \sin 2\phi\, H + \gamma r_{0}^{2} \sin \phi = 0. \label{1linearode}
\end{equation}
Let us take $\tilde{P}(\xi) = \textmd{exp}
\left(-\gamma r_{0}^{2} \int \sin 2 \phi(\xi)\, d\xi \right)$ as the integrating factor, multiply it to \eqref{1linearode} and integrate the result with respect to $\xi$, then we obtain the solution for $H$:
\begin{equation}
   H(\xi,\tau) = \frac{- \gamma r_{0}^{2} \int \tilde{P}(\xi) \sin \phi(\xi) \, d\xi - \zeta(\tau)}{\tilde{P}(\xi)} = \frac{\tilde{Q}(\xi) - \zeta(\tau)}{\tilde{P}(\xi)},
\end{equation}
where $-\zeta(\tau)$ is a constant of integration that depends on $\tau$ and $\tilde{Q}(\xi)$ denotes the term with the integral sign. Assuming that the displaced phase $\phi(\xi)$ is an invertible function, we can write $\xi = \xi(\phi)$ and dropping the tilde signs to indicate that $P = P(\phi)$ and $Q = Q(\phi)$. Consequently, $G$ is now written as follows:
\begin{equation}
  G(\phi,\tau) = \frac{P(\phi)}{Q(\phi) - \zeta(\tau)}. \label{AnsatzG}
\end{equation}
We will observe in the following subsections that by choosing three different functions of $\zeta(\tau)$ will lead to three different breather solutions of the NLS equation.

On the other hand, multiplying the real part by $\sin
\phi$ and the imaginary part by $\cos \phi$, subtracting one from
the other we obtain a nonlinear oscillator equation for $G$:
\begin{equation}
  \beta \partial_{\tau}^{2} G + (\phi'(\xi) + 2\gamma r_{0}^{2}\cos^{2}\phi)G - 3 \gamma r_{0}^{2} \cos \phi G^{2} + \gamma r_{0}^{2} G^{3} = 0.%
  \label{oscillator}
\end{equation}
In deriving the breather solutions, we will compare this equation with the second order differential equation for $G$ after choosing a particular function of $\zeta(\tau)$.

\begin{center}
{\bf 2.1 Soliton on Finite Background}
\end{center}

Firstly, we take $\zeta(\tau) = \cos(\nu \tau)$, where $\nu$ is a
modulation frequency. For the normalized quantity $\tilde{\nu} =
\nu/\left(r_{0}\sqrt{\frac{\gamma}{\beta}}\right)$ we consider $0
< \tilde{\nu} < \sqrt{2}$. The differential equation for $G$
becomes:
\begin{equation}
  \partial_{\tau}^{2}G = -\nu^{2} G + 3\nu^{2} \frac{Q}{P} G^{2} + 2\nu^{2}\frac{1 - Q^{2}}{P^{2}} G^{3}. \nonumber%
\end{equation}
By comparing with (\ref{oscillator}), the solution (\ref{AnsatzG})
is obtained with $P(\phi) = \tilde{\nu}^{2} Q(\phi)/\cos \phi$,
$Q^{2}(\phi) = 2 \cos^{2}\phi /(2 \cos^{2}\phi - \tilde{\nu}^{2})$
and the displaced phase satisfies $\tan \phi(\xi) =
-\frac{\tilde{\sigma}} {\tilde{\nu}^{2}} \tanh(\sigma \xi)$, where
$\tilde{\sigma} = \tilde{\nu} \sqrt{2 - \tilde{\nu}^{2}}$ and
$\sigma = \gamma r_{0}^{2} \tilde{\sigma}$. The corresponding
solution of the NLS equation (\ref{NLSsolution}) with extremum at
$(\xi,\tau) = (0,0)$ can then be written after some manipulations as:%
\begin{equation}
  A(\xi,\tau) = A_{0}(\xi) \cdot \left(\frac{\tilde{\nu}^{2} \cosh(\sigma \xi) - i \tilde{\sigma} \sinh(\sigma \xi)}
                {\cosh(\sigma \xi) - \sqrt{1 - \frac{1}{2} \tilde{\nu}^{2}} \cos(\nu \tau)} - 1 \right). \nonumber%
\end{equation}

\begin{figure}[h]
  \begin{center}
    \subfigure[]{%\epsfig{file=SFB3D.eps,      width=0.3\textwidth}}
    \includegraphics[bb = 55 234 561 598,width=0.3\textwidth]{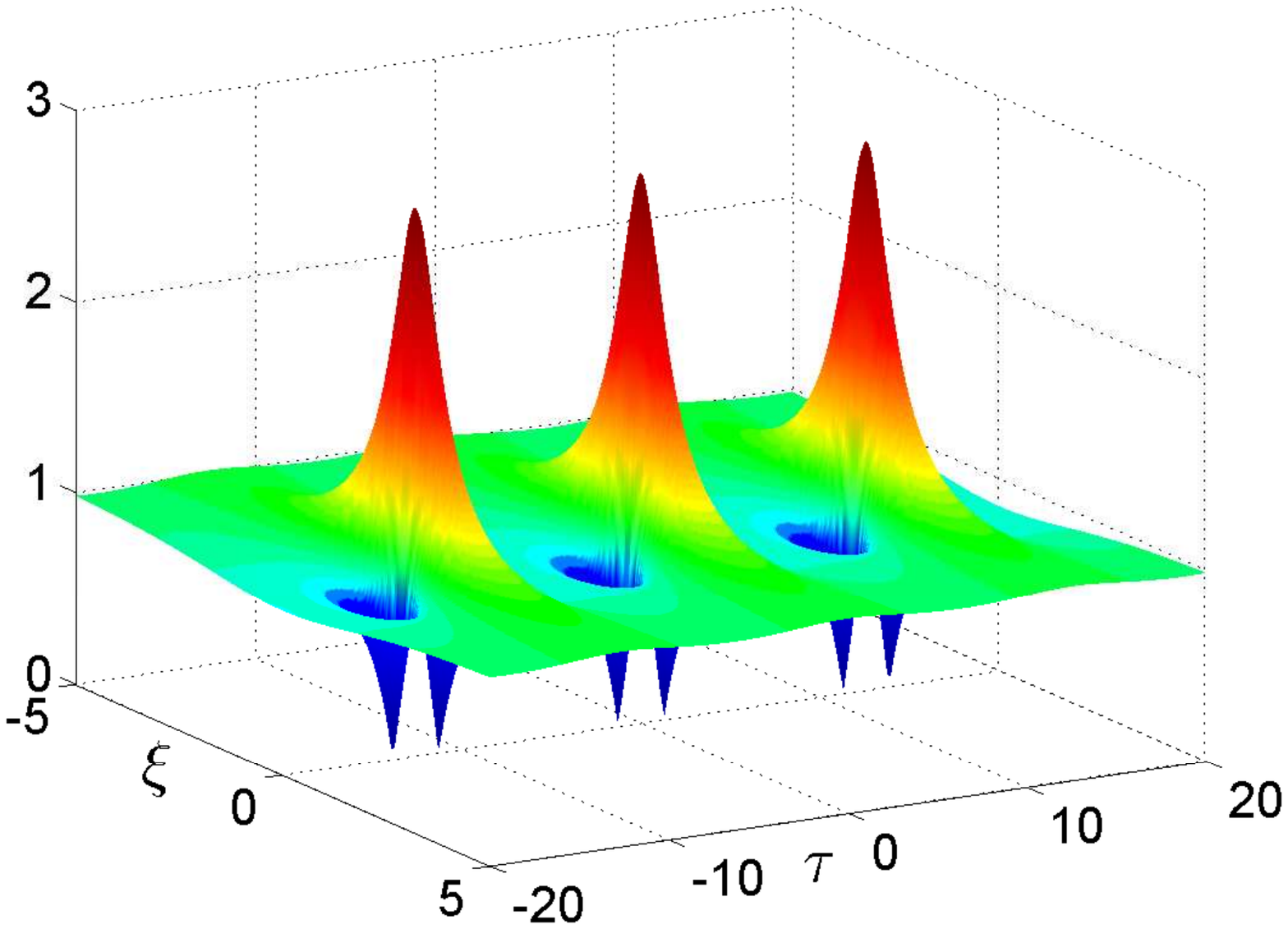}}
    \hspace{0.3cm}
    \subfigure[]{%\epsfig{file=Ma_3D.eps,      width=0.3\textwidth}}
    \includegraphics[bb = 22 181 555 615,width=0.3\textwidth]{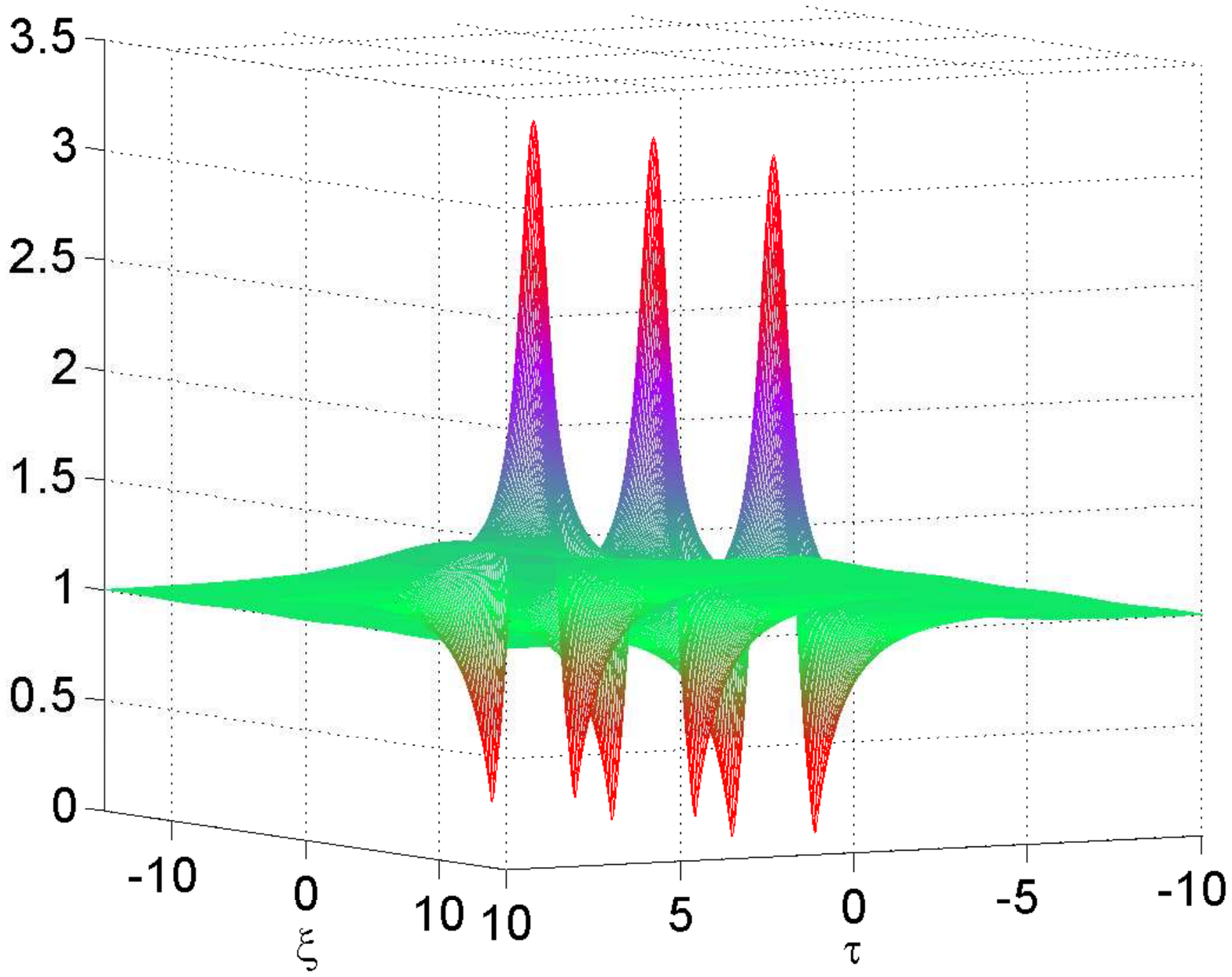}}
    \hspace{0.3cm}
    \subfigure[]{%\epsfig{file=Rational_3D.eps,width=0.3\textwidth}}\\
    \includegraphics[bb = 22 182 555 615,width=0.3\textwidth]{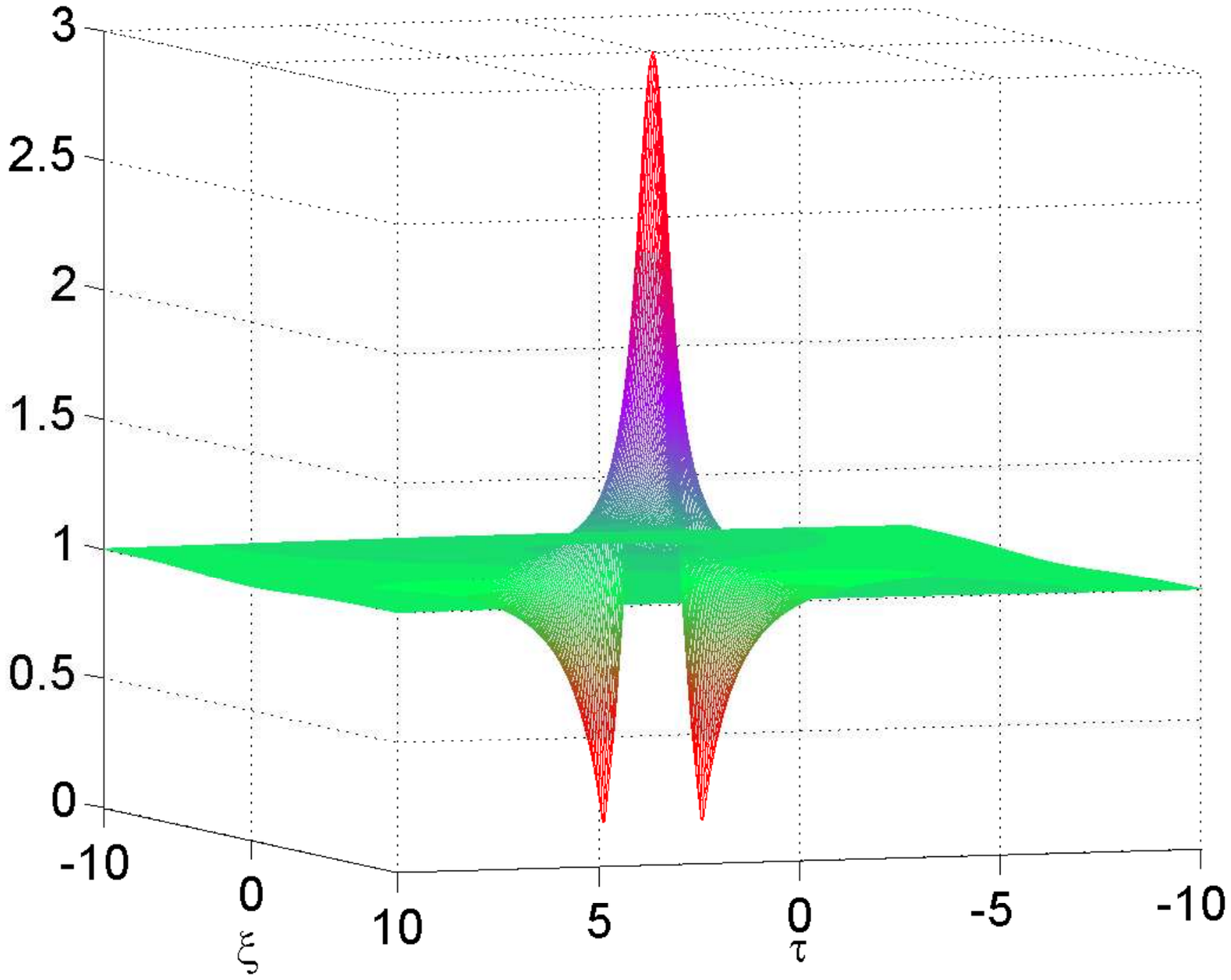}}\\
%    \includegraphics[width=0.3\textwidth]{Rational_3D.pdf}}\\
    %(a) \hspace{5cm} (b) \hspace{5cm} (c)\\
    \caption{\footnotesize Plots of the absolute value of the SFB breather
    for $\tilde{\nu} = 1/2$ (left), the Ma breather for $\tilde{\mu} = 0.4713$ (center) and
    the rational breather (right), for $r_{0} = 1$ in all cases.
    For illustration purposes, the axis are scaled corresponding to $\beta = 1 = \gamma$.} \label{3D}
\end{center}
\end{figure}

This solution was found by Akhmediev et. al. (1987). They
suggested a method of obtaining exact solutions of the NLS
equation which is based on the substitution connecting the real
and imaginary parts of the solution by a linear relationship with
coefficients depending only on time. The method consists in
constructing a certain system of ordinary differential equations,
the solutions of which determine the solutions of the NLS
equation. However, the explicit expression of the SFB already
appeared in the previous papers (Akhmediev et. al. 1985; Akhmediev
and Korneev, 1986). The SFB is also derived by Ablowitz and
Herbst (1990) using Hirota's method (Hirota, 1976) and for
$\tilde{\nu} = 1$ by Osborne et. al. (2000) using the inverse
scattering technique. The SFB is the single homoclinic orbit to
the cw as a fixed point of the NLS equation (Ablowitz and Herbst,
1990; Li and McLaughlin, 1994, 1997; Calini and Schober, 2002).
The asymptotic behaviour for $\xi \rightarrow -\infty$ of the SFB
describes the linear instability of the cw solution for side-band
perturbation (Akhmediev and Korneev, 1986). This modulational
instability is known as the Benjamin-Feir instability in the
context of water waves (Benjamin and Feir, 1967). A plot of the
absolute value of this complex amplitude solution for $r_{0} = 1$
and $\tilde{\nu} = 1/2$ is given in Figure \ref{3D}(a).

\begin{center}
{\bf 2.2 Ma breather}
\end{center}

Secondly, we take $\zeta(\tau) = \cosh(\mu \tau)$, where $\mu =
r_{0} \sqrt{\frac{\gamma}{\beta}} \tilde{\mu}$. The differential
equation for $G$ becomes:
\begin{equation}
  \partial_{\tau}^{2}G = \mu^{2} G - 3\mu^{2}\frac{Q}{P} G^{2} - 2\mu^{2} \frac{1 - Q^{2}}{P^{2}}
  G^{3}. \nonumber
\end{equation}
Comparing again with (\ref{oscillator}), the solution
(\ref{AnsatzG}) is obtained with $P(\phi) = -\tilde{\mu}^{2}
Q(\phi)/\cos \phi$, $Q^{2}(\phi) = 2 \cos^{2}\phi/(2 \cos^{2}\phi
+ \tilde{\mu}^{2})$ and the displaced phase satisfies $\tan
\phi(\xi) = - \frac{\tilde{\rho}}{\tilde{\mu}^{2}} \tan(\rho
\xi)$, where $\tilde{\rho} = \tilde{\mu} \sqrt{2 +
\tilde{\mu}^{2}}$ and $\rho = \gamma r_{0}^{2} \tilde{\rho}$. The
corresponding solution of the NLS (\ref{NLSsolution}) can then be
written after some manipulations as:
\begin{equation}
  A(\xi,\tau) = A_{0}(\xi) \cdot \left(\frac{-\tilde{\mu}^{2} \cos(\rho \xi) + i \tilde{\rho} \sin(\rho \xi)}
                {\cos(\rho \xi) - \sqrt{1 + \frac{1}{2} \tilde{\mu}^{2}} \cosh(\mu \tau)} - 1 \right). \nonumber%
\end{equation}
We call this solution Ma solution or Ma breather since it has been
found by Ma (1979) using the inverse scattering technique of
(Gardner et. al., 1967). The same technique is also used by
Osborne et. al. (2000) to arrive at this solution for
$\tilde{\mu} = \sqrt{2}$. A plot of the absolute value of the Ma
breather for $r_{0} = 1$ and $\tilde{\mu} = 0.4713$ is given in
Figure \ref{3D}(b). This choice of this value of $\tilde{\mu}$ is
just for our convenience, which we will use again for the
corresponding physical wave field.

\begin{center}
{\bf 2.3 Rational breather}
\end{center}

Thirdly, we take $\zeta(\tau) = 1 - \frac{1}{2} \nu^{2} \tau^{2}$.
Similarly, the same solution can also be obtained by substituting
$\zeta(\tau) = 1 + \frac{1}{2} \mu^{2} \tau^{2}$. As a
consequence, the differential equation for $G$ now becomes:
\begin{equation}
  \partial_{\tau}^{2}G = \frac{3\nu^{2}}{P} G^{2} - 4\nu^{2} \frac{Q - 1}{P^{2}} G^{3}. \nonumber%
\end{equation}
Comparing again with (\ref{oscillator}), we have $P(\phi) =
\tilde{\nu}^{2}/\cos \phi$, $Q(\phi) = 1 +
P^{2}/(4\tilde{\nu}^{2})$ and the displaced phase satisfies $\tan
\phi(\xi) = - 2\gamma r_{0}^{2} \xi$. Substituting into the Ansatz
(\ref{AnsatzG}), we obtain
\begin{equation}
  A(\xi,\tau) = A_{0}(\xi) \cdot \left(\frac{4 (1 - 2 i \gamma r_{0}^{2} \xi)}
                {1 + 4 (\gamma r_{0}^{2} \xi)^{2} + 2 \frac{\gamma}{\beta} r_{0}^{2} \tau^{2}} - 1 \right). \nonumber%
\end{equation}
This rational solution is clearly different from the two solutions
above, is derived by Peregrine (1983) as a limiting case from the
Ma breather. He took the analytic expression and performed a
double Taylor series expansion about the amplitude peak (which
occur at $\xi = 0 = \tau$) to arrive at his solution. Since this
solution is confined in both space and time, some authors
(Henderson et. al., 1999) also called it the `isolated Ma soliton'
and other authors (Akhmediev and Ankiewicz, 1997) called it the
`rational solution'. Because of its soliton-like feature in $\xi$,
(Nakamura and Hirota, 1985) called it as an `explode-decay
solitary wave'. In this letter, we call it the `rational
breather'. A plot of the absolute value of the rational breather
for $r_{0} = 1$ is given in Figure \ref{3D}(c).

\vspace{1.5cc}

\begin{center}
{\bf 3. RELATION BETWEEN THE NLS BREATHER SOLUTIONS}%
\end{center}

As shown by Dysthe and Trulsen (1999), there is a relationship
among the breather solutions of the NLS equation. Different
parameters are used for explicit expressions of the SFB and the Ma
breather in Dysthe and Trulsen (1999) as well as in Grimshaw et.
al. (2001). Introducing new parameters $\varphi$ and $\vartheta
\in \mathbb{R}$ by the following relations: $\tilde{\nu} =
\sqrt{2} \sin \varphi$, $\tilde{\sigma} = \sin 2\varphi$,
$\tilde{\mu} = \sqrt{2} \sinh \vartheta$ and $\tilde{\rho} = \sinh
2\vartheta$, we can write our expressions similar to the ones in
these papers.
\begin{figure}[h!]
\begin{center}
  \includegraphics[bb = 73 597 445 815,width=0.4\textwidth]{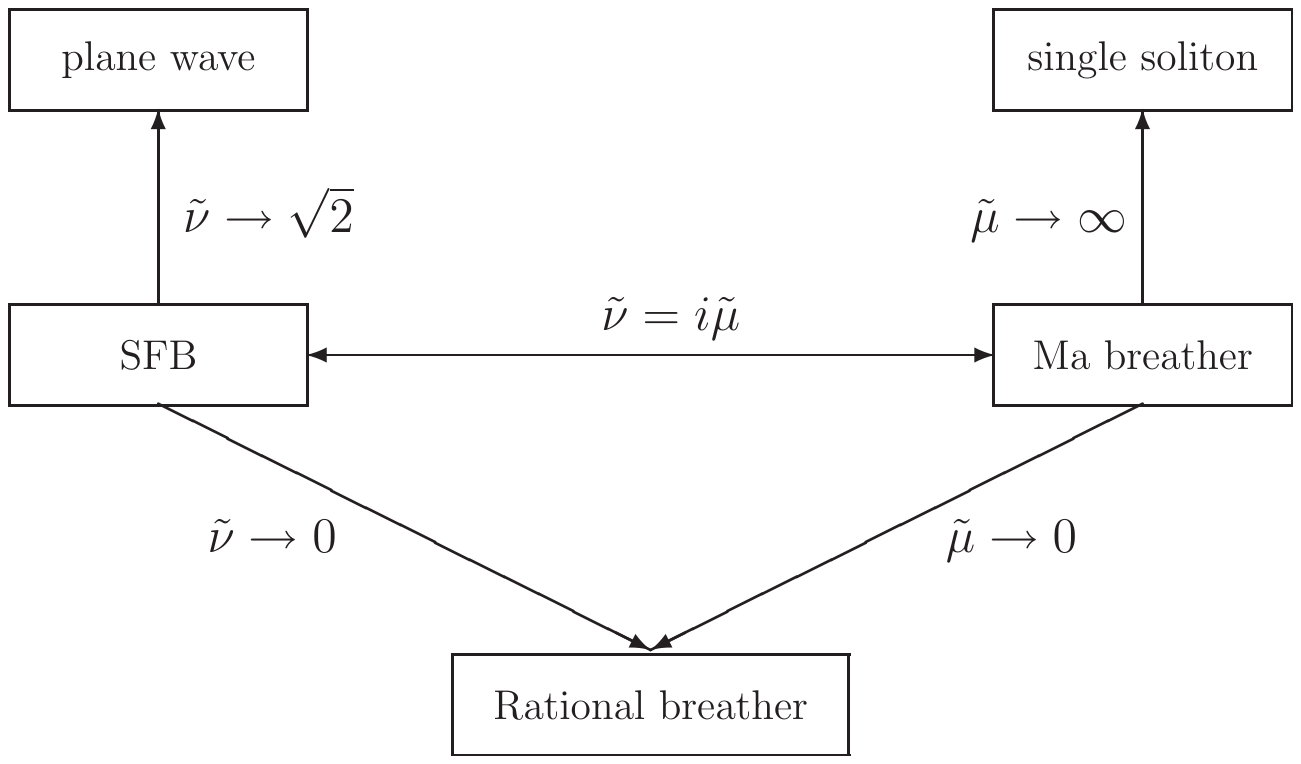}
  \caption{\footnotesize The schematic diagram for the derivation of
  the Ma breather and the rational breather from the SFB.}
  \label{segitiga}%
\end{center}
\end{figure}

The SFB becomes the Ma breather if we substitute $\nu = i \mu$ and
it becomes the rational breather if $\nu \rightarrow 0$.
Similarly, the Ma breather becomes the rational breather for $\mu
\rightarrow 0$. Interestingly, the Ma breather becomes the
one-soliton solution for $\mu \rightarrow \infty$ (Akhmediev and
Ankiewicz, 1997). Figure \ref{segitiga} explains the schematic
diagram of these derivations. These relations can also be seen from the expression in the
previous section. For the SFB, we take $\zeta(\tau) = \cos (\nu
\tau)$, which leads to the Ma breather if we substitute $\nu = i
\mu$, so that $\zeta(\tau) = \cos (i \mu \tau) = \cosh (\mu
\tau)$. Taking either $\nu$ or $\mu$ $\rightarrow 0$, we get
$\zeta(\tau) = 1 - \frac{1}{2} \nu^{2} \tau^{2}$ or $\zeta(\tau) =
1 + \frac{1}{2} \mu^{2} \tau^{2}$, and both will lead to the
rational breather.
\begin{figure}[h]
  \begin{center}
  \vspace*{0.5cm}
  \includegraphics[bb = 62 223 550 572,width=0.35\textwidth]{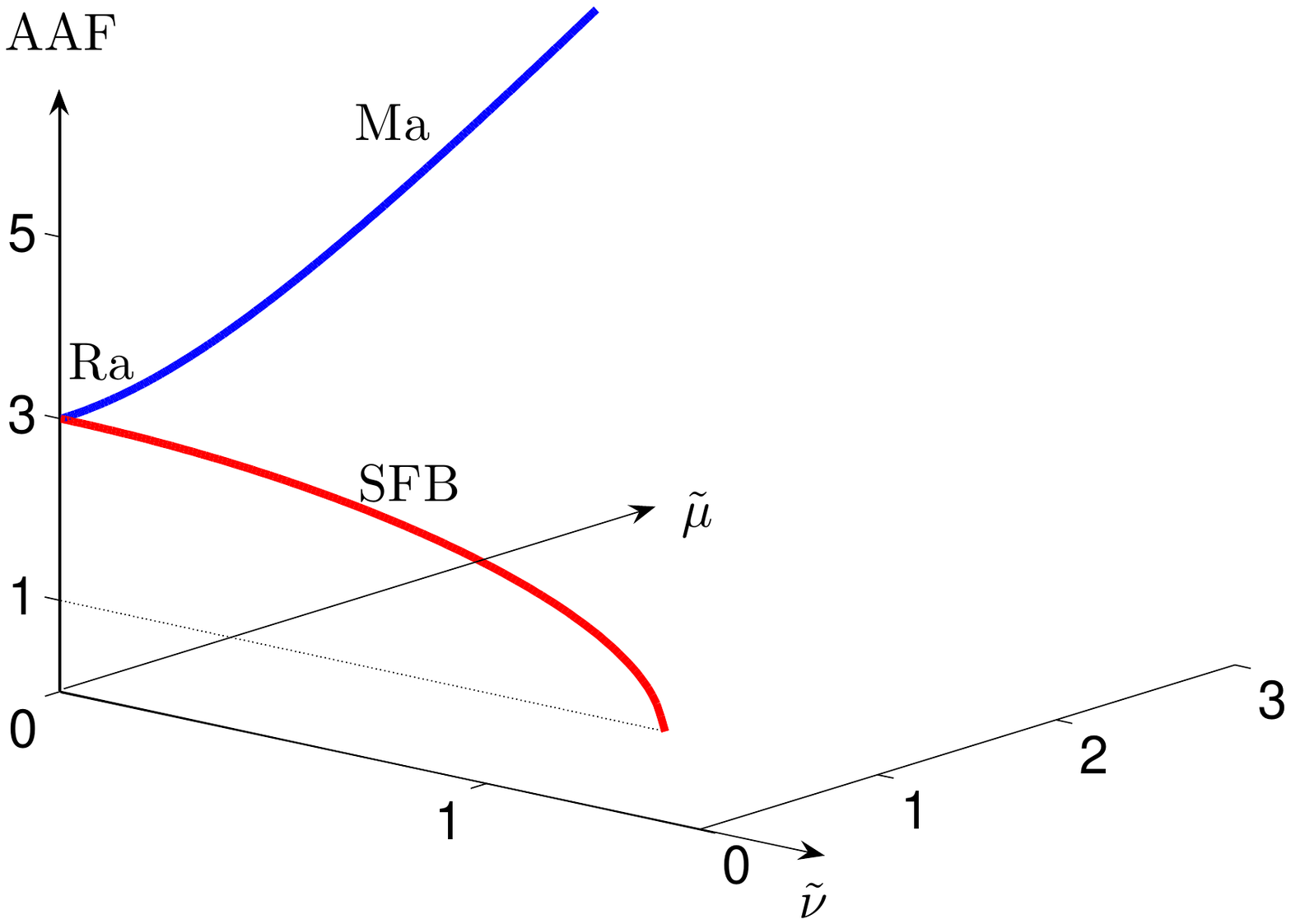}
  \caption{\footnotesize Plot of the amplitude amplification factor for
  the SFB and the Ma breather. The meeting point of the two curves
  is the AAF of the rational breather.}
  \label{AAF_plot}
  \end{center}
\end{figure}

The relations have also consequence for the amplitude
amplification factor (AAF), defined as the ratio between the
maximum amplitude and the value of its background. The expressions
for the breather solutions as given in this letter were shifted
such that the maximum amplitude is at $(\xi,\tau) = (0,0)$ and the
value of the background is $r_{0}$. For the SFB, the amplification
is given by $\textmd{AAF}_{\textmd{\tiny S}}(\tilde{\nu}) = 1 +
\sqrt{4 - 2\tilde{\nu}^{2}}$. For $0 < \tilde{\nu} < \sqrt{2}$,
the amplification is bounded and are $1 <
\textmd{AAF}_{\textmd{\tiny S}}(\tilde{\nu}) < 3$. The AAF for the
Ma breather is given by $\textmd{AAF}_{\textmd{\tiny
Ma}}(\tilde{\mu}) = 1 + \sqrt{4 + 2\tilde{\mu}^{2}}$. Hence, for
$\tilde{\mu} > 0$, we have $\textmd{AAF}_{\textmd{\tiny Ma}} > 3$.
The AAF for the rational breather is exactly $3$, which follows by
letting the either $\tilde{\nu}$ or $\tilde{\mu}$ go to zero
\begin{equation}
\textmd{AAF}_{\textmd{\tiny Ra}} =  \lim_{\tilde{\nu} \rightarrow
0} \textmd{AAF}_{\textmd{\tiny S}} = 3 = \lim_{\tilde{\mu}
\rightarrow 0} \textmd{AAF}_{\textmd{\tiny Ma}}.
\end{equation}
In Onorato et. al. (2001), the AAF for the SFB is given as a
function of the wave steepness and the number of waves under the
modulation. The plot of the AAF for all the three breather
solutions is given in Figure \ref{AAF_plot}.
For $\tilde{\mu} \rightarrow \infty$ in the Ma breather, the
single-soliton solution is obtained (Akhmediev and Ankiewicz,
1997).

\begin{center}
{\bf 4. EVOLUTION IN THE ARGAND DIAGRAM}
\end{center}

In this section we present the evolution of the breather solutions
in a complex plane. The real and the imaginary axes of this plane
are denoted by the real and the imaginary part of the breather
solutions after removing the plane-wave contribution,
respectively. In this paper we refer this plane as the Argand
diagram. Since the breather solutions depend on the space variable
$\xi$ and the time variable $\tau$, the evolution curves are
presented by parameterizing one variable and plotting for the
other at different values. We will observe that the point $(-1,0)$
plays a significant role with respect to all evolution curves,
whether they are parameterized in space or in time. A study of
this evolution curves for understanding experimental results on
freak wave generation using the SFB solution has been presented
recently (Karjanto, 2006).
\begin{figure}[h]
  \begin{center}
    \subfigure[]{%\epsfig{file = 1ArgSFBTauNu1sqrt2,width = 0.3\textwidth}}
    \includegraphics[bb = 97 201 570 613,width=0.3\textwidth]{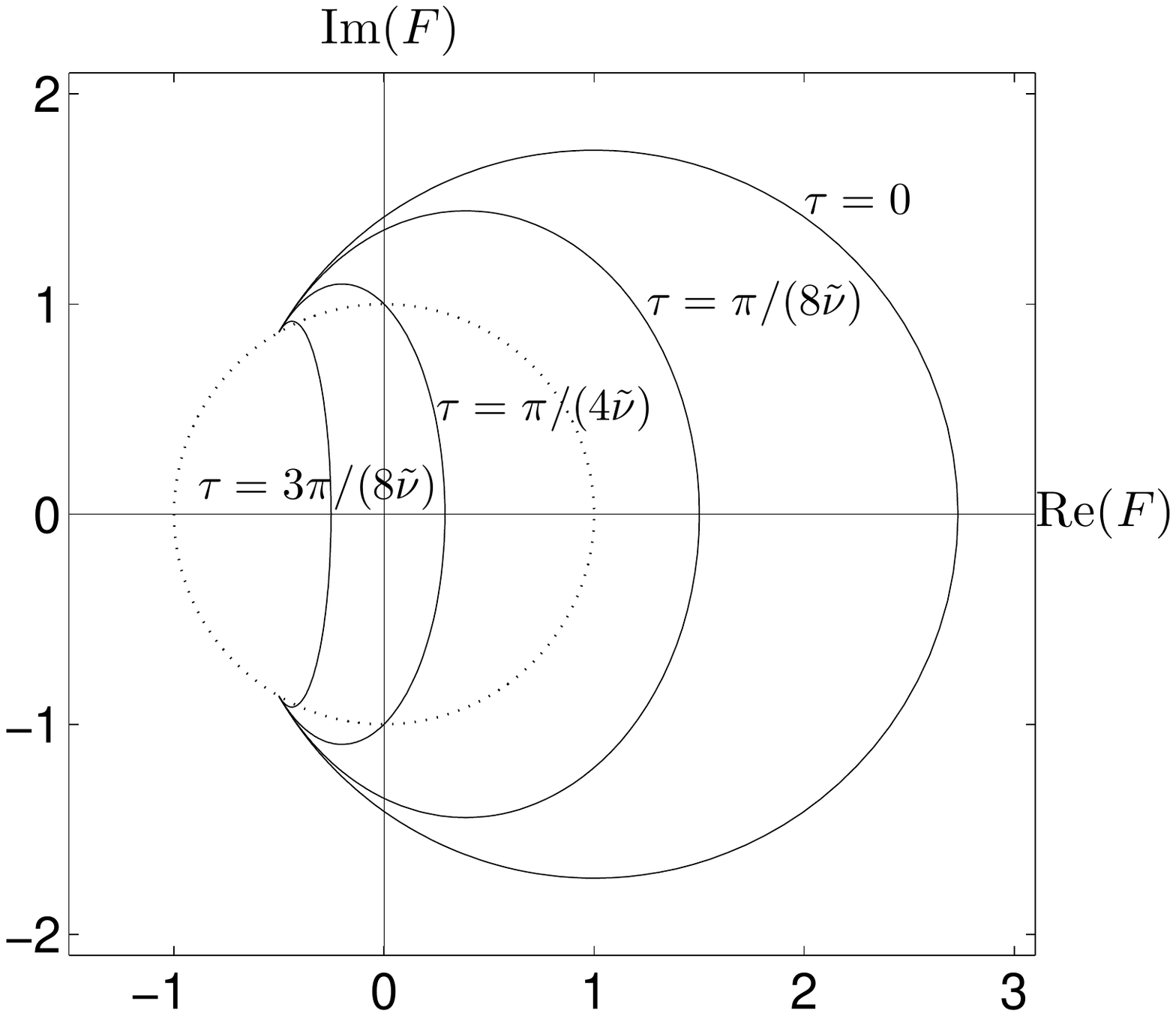}}
    \hspace{0.5cm}
    \subfigure[]{%\epsfig{file = 1ArgSFBXiNu1sqrt2, width = 0.3\textwidth}}
    \includegraphics[bb = 97 201 578 613,width=0.3\textwidth]{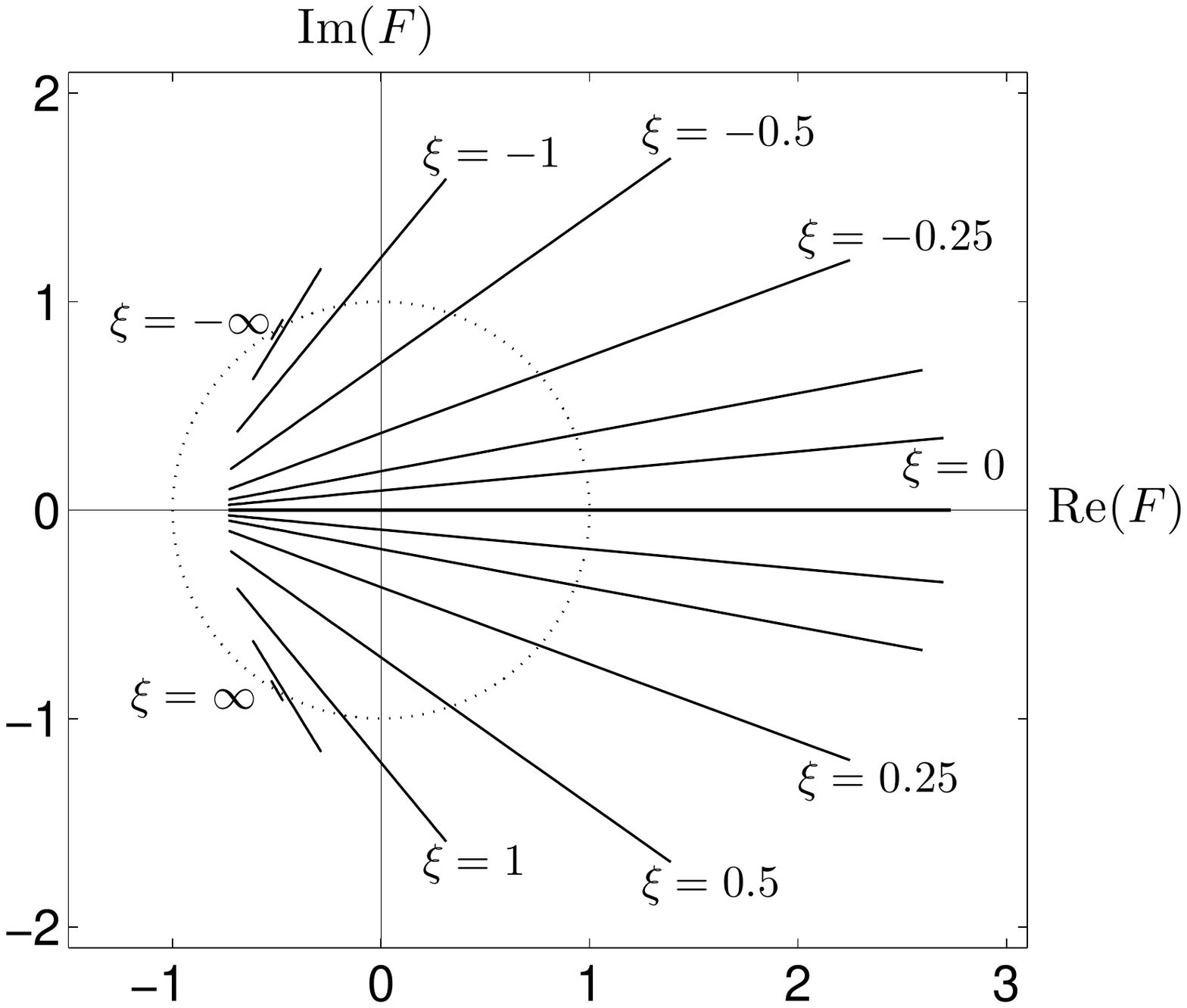}}
    \caption{\footnotesize The evolution of the SFB in the Argand diagram for $\tilde{\nu} = 1/\sqrt{2}$.}
    \label{ArgSFB}
  \end{center}
\end{figure}

\begin{figure}[h!]
  \begin{center}
    \hspace*{-1.2cm}
    \subfigure[]{%\epsfig{file = 1ArgRaTau,width = 0.3\textwidth}}
    \includegraphics[bb = 81 179 496 583,width=0.3\textwidth]{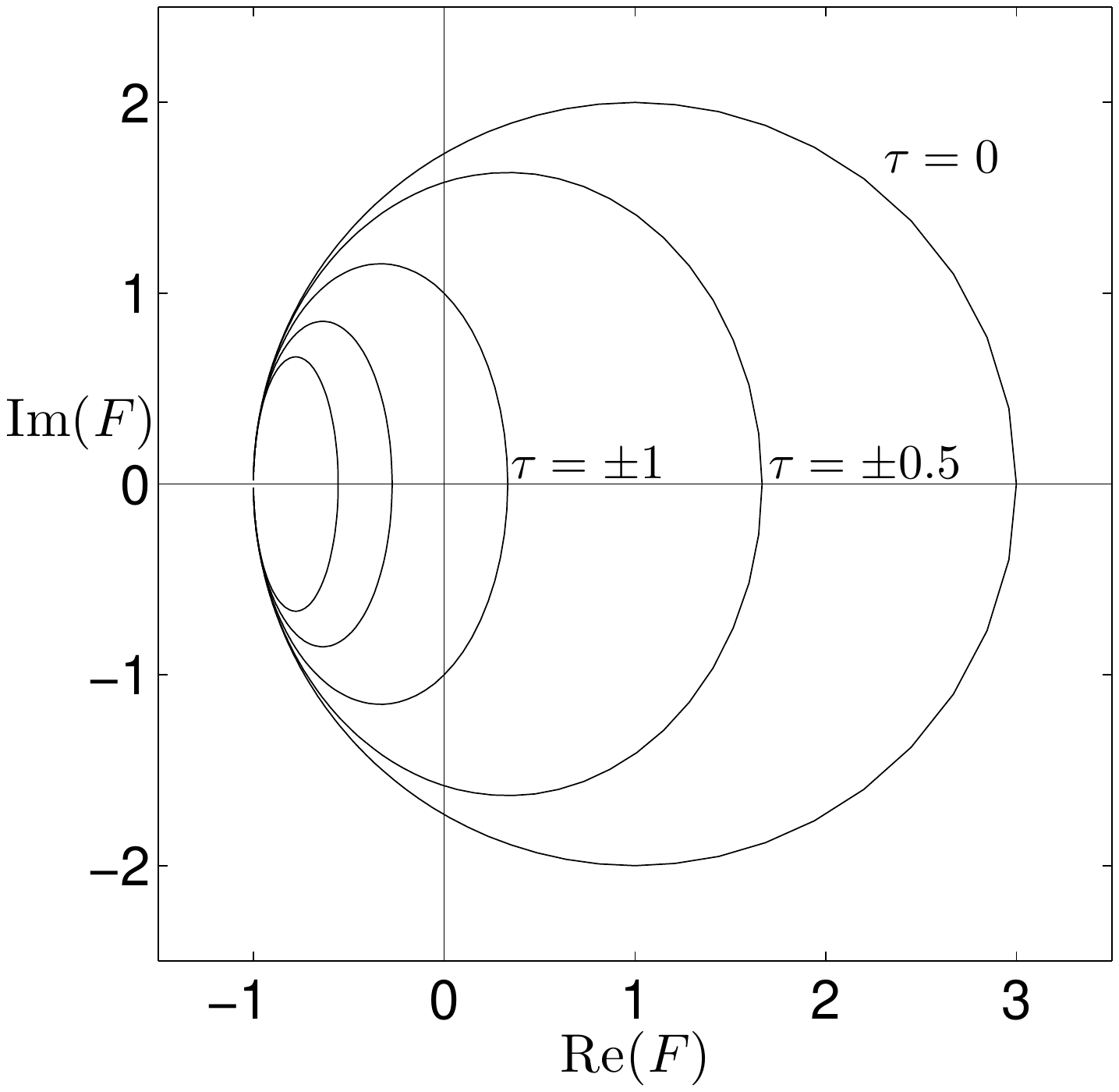}}
    \hspace{0.5cm}
    \subfigure[]{%\epsfig{file = 1ArgRaXi, width = 0.3\textwidth}}
    \includegraphics[bb = 78 179 496 583,width=0.3\textwidth]{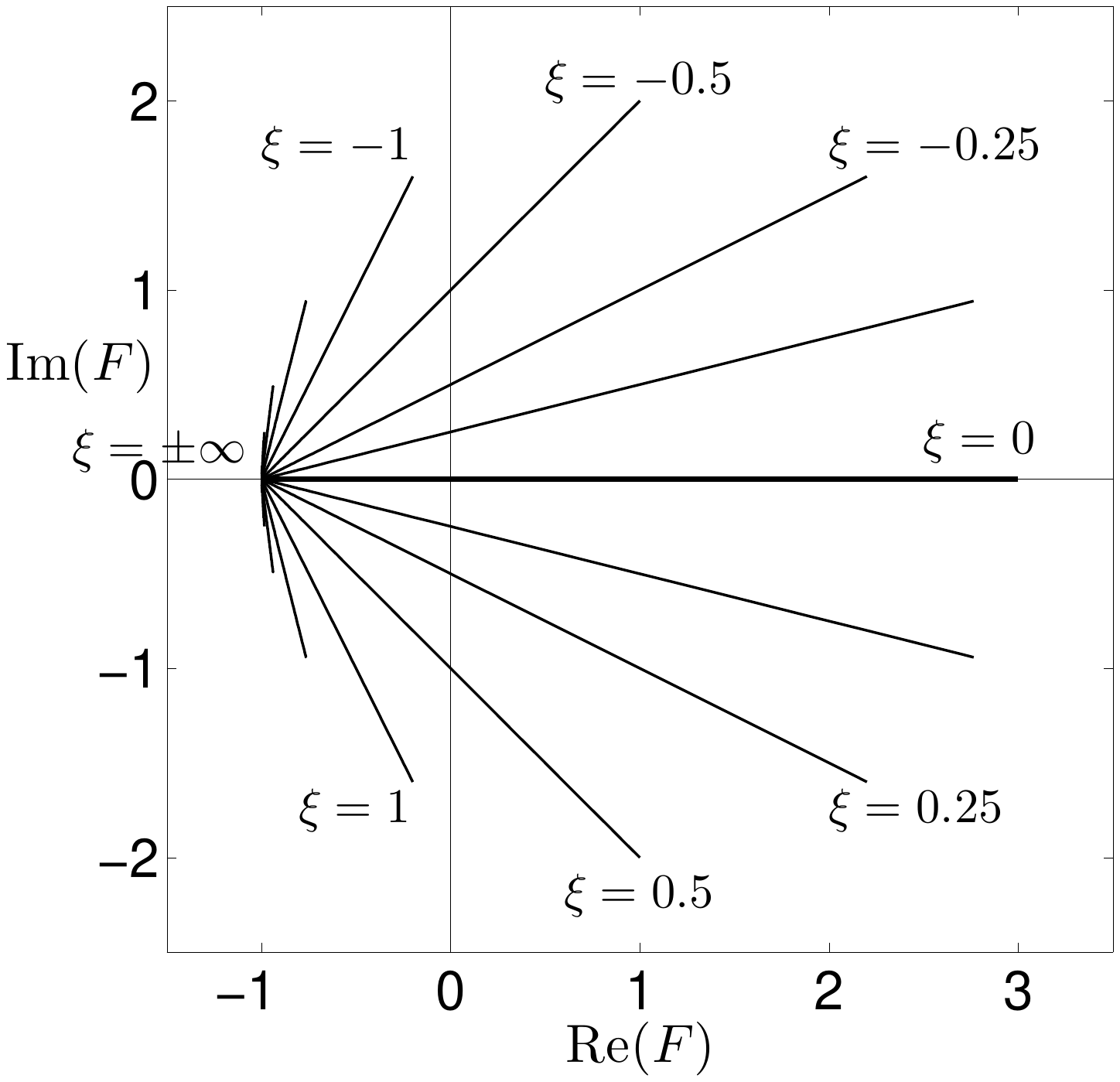}}
    \caption{\footnotesize The evolution of the rational breather in the Argand diagram.}
    \label{ArgRa}
  \end{center}
\end{figure}

Figure \ref{ArgSFB} shows the evolution curves of the SFB for
$\tilde{\nu} = 1/\sqrt{2}$. The parameterized plot in position for
different times is given by Figure \ref{ArgSFB}(a). For this
particular value of modulation frequency $\tilde{\nu}$, all curves start
from a point in the second quadrant, which corresponds to $\xi =
-\infty$, and end to a point in the third quadrant, which
corresponds to $\xi = \infty$. During one modulation period $0
\leq \tau < 2\pi/\tilde{\nu}$, we observe that the curve reaches its
maximum value for $\tau = 0$ when the Im($F$) vanishes and it
reaches a minimum value for $\tau = \pi/\tilde{\nu}$. The dotted circle
indicates the plane wave solution that has been removed. When we allow
the modulation frequency $\tilde{\nu} \rightarrow 0$, both the initial and
the final points will coincide at $(-1,0)$, as shown by the
rational breather case in Figure \ref{ArgRa}(a). The parameterized
plot in time for different positions is given by Figure
\ref{ArgSFB}(b). All curves are straight lines and they are passed
twice during one modulation period. For $\xi \rightarrow \pm
\infty$, the lines shrink to a point in the third and the second
quadrants, respectively. The longest line is reached at $\xi = 0$
when it lies on the real axis. Notice that $(-1,0)$ becomes a
center for all the lines. Moreover, for $\tilde{\nu} \rightarrow 0$, the
left-hand side of all the lines will meet at $(-1,0)$, as shown by
the rational breather case in Figure \ref{ArgRa}(b).
\begin{figure}[h]
  \begin{center}
    \subfigure[]{%\epsfig{file = 1ArgMaTauMu2,width = 0.3\textwidth}}
    \includegraphics[bb = 114 201 557 609,width=0.3\textwidth]{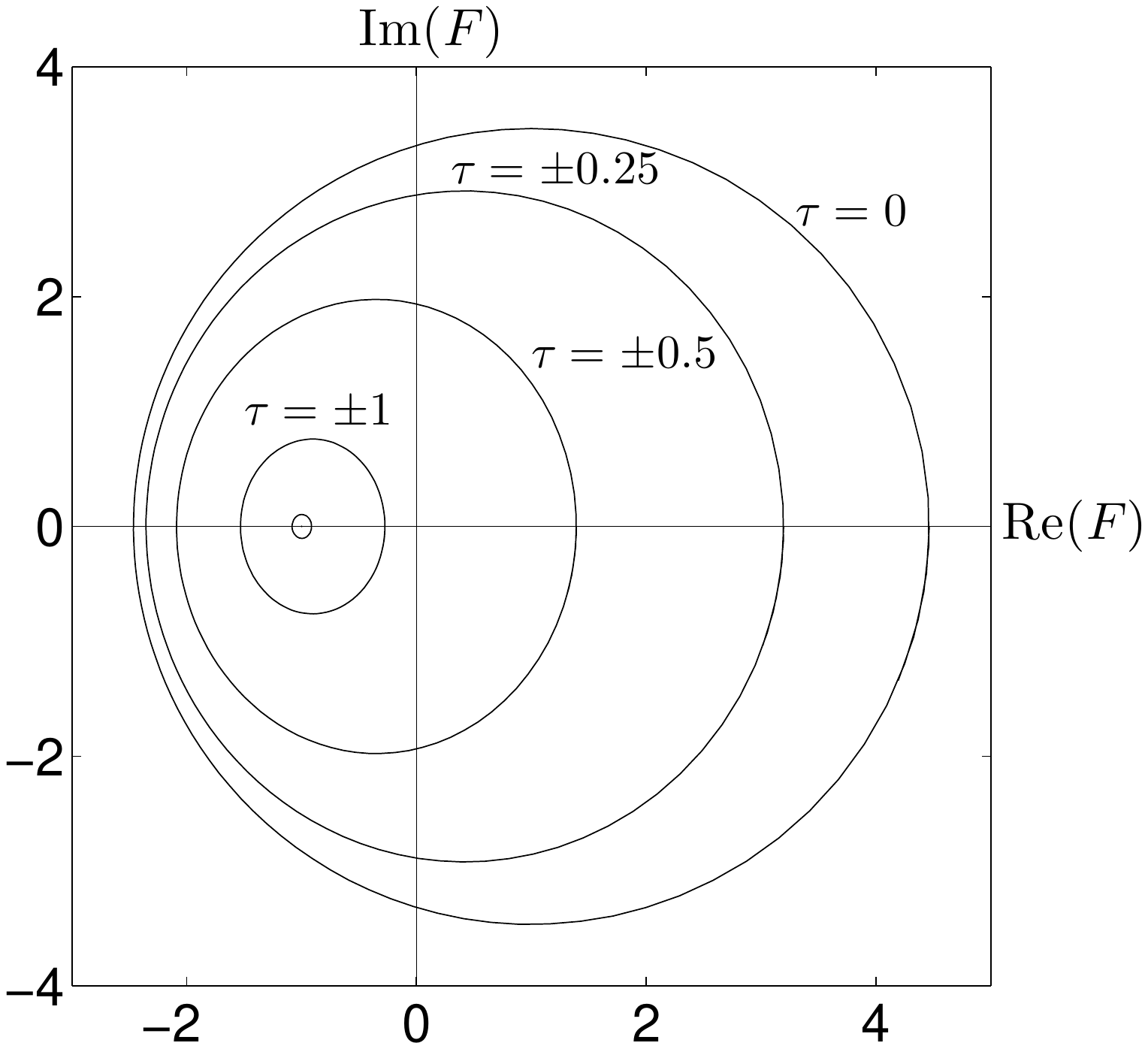}}
    \hspace{0.5cm}
    \subfigure[]{%\epsfig{file = 1ArgMaXiMu2, width = 0.3\textwidth}}
    \includegraphics[bb = 114 201 557 609,width=0.3\textwidth]{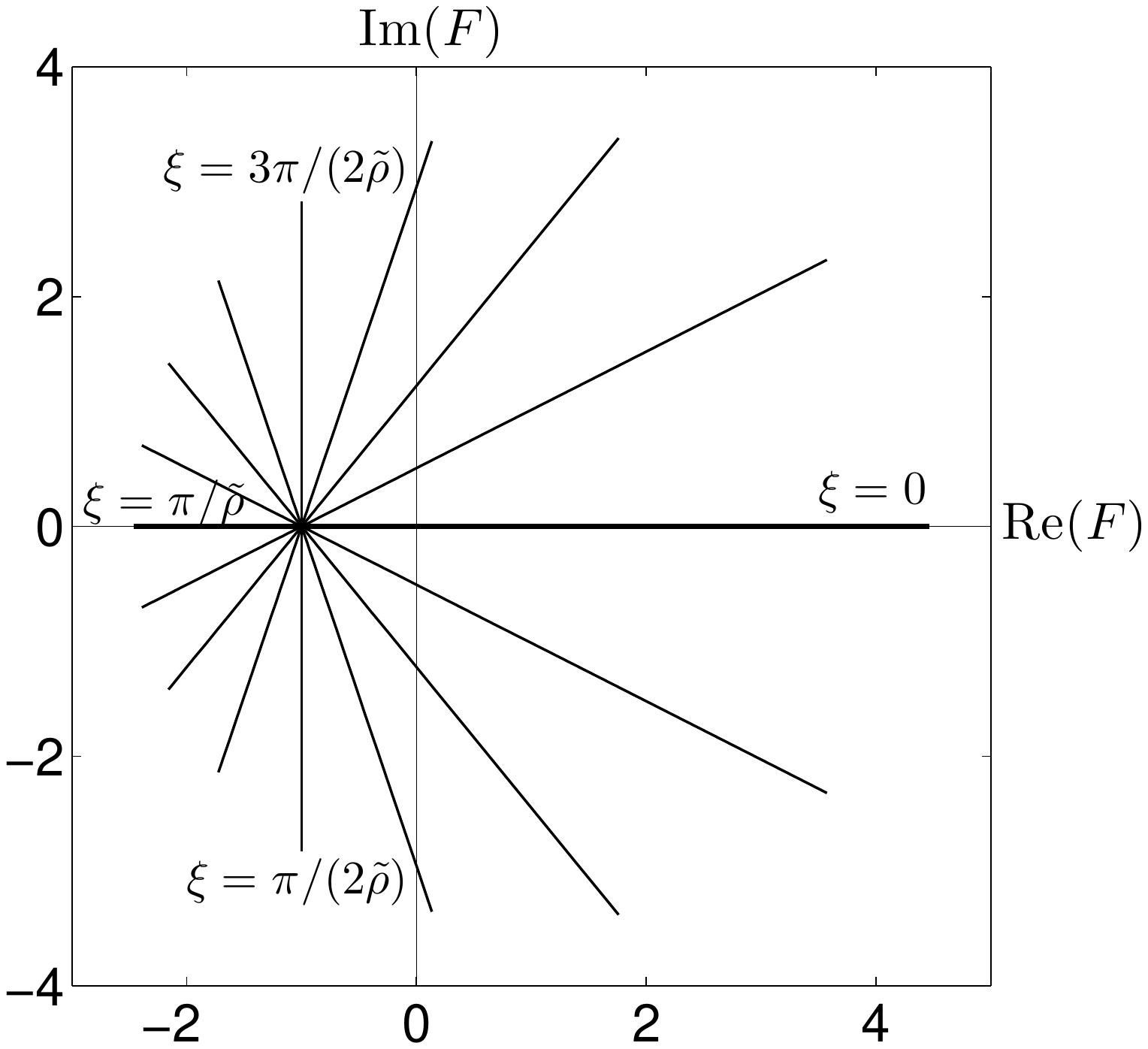}}
    \caption{\footnotesize The evolution of the Ma breather in the Argand diagram for $\tilde{\mu} = 2$.}
    \label{ArgMa}
  \end{center}
\end{figure}

Figure \ref{ArgMa} shows the evolution curves of the Ma breather
for $\tilde{\mu} = 2$. The parameterized plot in position for
different times is given by Figure \ref{ArgMa}(a). All the curves
are elliptical, which indicate that they are $\xi$-periodic. The
largest curve occurs at $\tau = 0$, which has a circular form. As
the time moves forward and backward, the ellipses are getting
smaller until eventually shrink to a point at $(-1,0)$ for $\tau =
\pm \infty$. By letting $\tilde{\mu} \rightarrow 0$, the left-hand
side of the ellipses at the real axis moves accordingly toward
$(-1,0)$, and eventually the pattern becomes the one as shown by
the rational breather in Figure \ref{ArgRa}(a). The parameterized
plot in time for different positions is given by Figure
\ref{ArgMa}(b). Similar to the previous case, all the curves are
straight lines and we notice clearly that they are centered at
$(-1,0)$. Different from the SFB solution case where the lines are
at the right-hand side of $(-1,0)$, for the Ma breather case the
point radiates lines to all different direction depending on the
position. For one periodic position $0 \leq \xi <
2\pi/\tilde{\rho}$, the longest line occurs at $\xi = 0$ and the
shortest one occurs at $\xi = \pi/\tilde{\rho}$, being both of
them lay on the real axis. When $\tilde{\mu} \rightarrow 0$, the
lines which are on the left-hand side of $(-1,0)$ will get shorter
until eventually become the rational breather case as shown in
Figure~\ref{ArgRa}(b). Figure \ref{ArgRa} shows the evolution curves
of the rational breather, which show the limiting cases for both
$\tilde{\nu} \rightarrow 0$ and $\tilde{\mu} \rightarrow 0$.

\vspace{1.5cc}

\begin{center}
{\bf 5. PHYSICAL WAVE FIELDS}
\end{center}

In this section, we show applications in surface water waves.
Since the NLS equation is an envelope equation, the corresponding
physical wave fields $\eta(x,t)$ are wave packets. Considering
only first order contributions, the physical wave field is given
by
\begin{equation}
  \eta(x,t) = A(\xi,\tau) e^{i(k_{0}x - \omega_{0}t)} + \textmd{c\,c}
\end{equation}
where c\,c denotes complex conjugate of the preceding term. The
complex amplitude $A$ is described in a moving frame of reference
with $\xi = x$ and $\tau = t - x/V_{0}$. The wavenumber $k_{0}$
and the frequency $\omega_{0}$ are related by the linear
dispersion relation. For surface water waves, the normalized form
of the dispersion relation is given by $\omega = \Omega(k) = k
\sqrt{(\tanh k)/k}$, the group velocity $V_{0} = \Omega'(k_{0})$.

\begin{figure}[h]
  \begin{center}
    \subfigure[]{%\epsfig{file=SFBphysical.eps,     width=0.25\textwidth}}
    \includegraphics[bb = 57 188 551 587,width=0.25\textwidth]{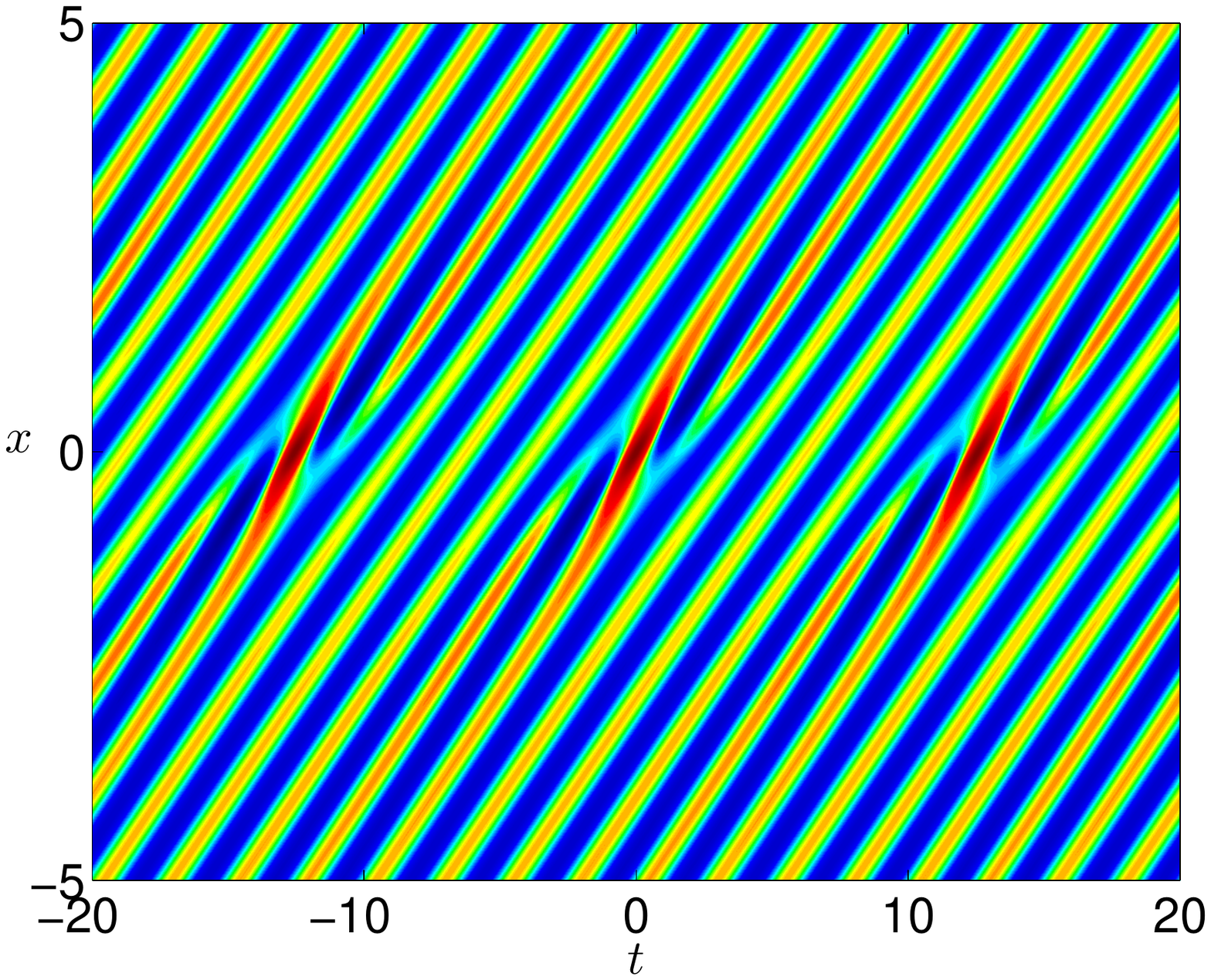}}
    \hspace{0.25cm}
    \subfigure[]{%\epsfig{file=Maphysical.eps,      width=0.25\textwidth}}
    \includegraphics[bb = 46 211 542 599,width=0.25\textwidth]{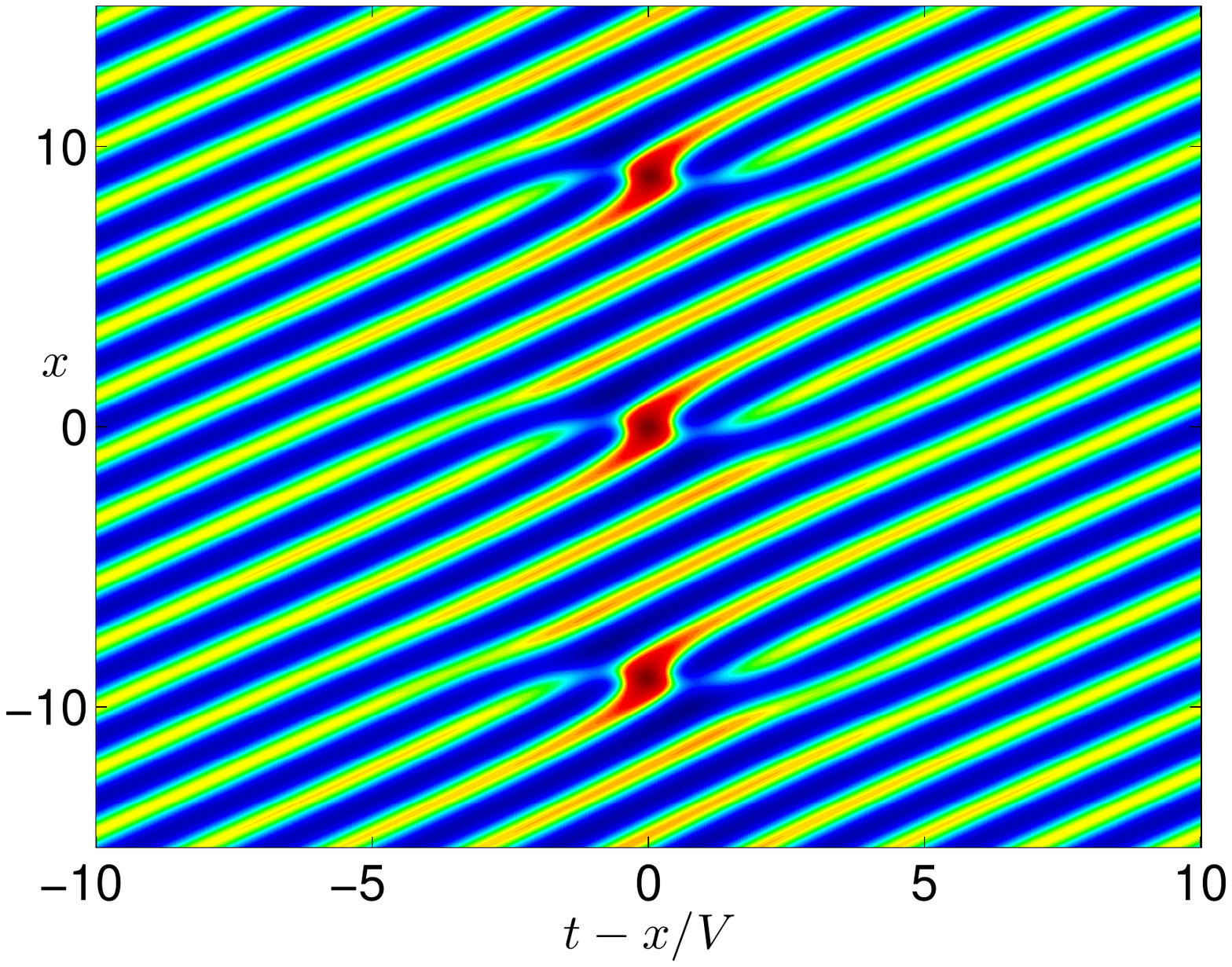}}
    \hspace{0.25cm}
    \subfigure[]{%\epsfig{file=Rationalphysical.eps,width=0.25\textwidth}}
    \includegraphics[bb = 48 218 542 605,width=0.25\textwidth]{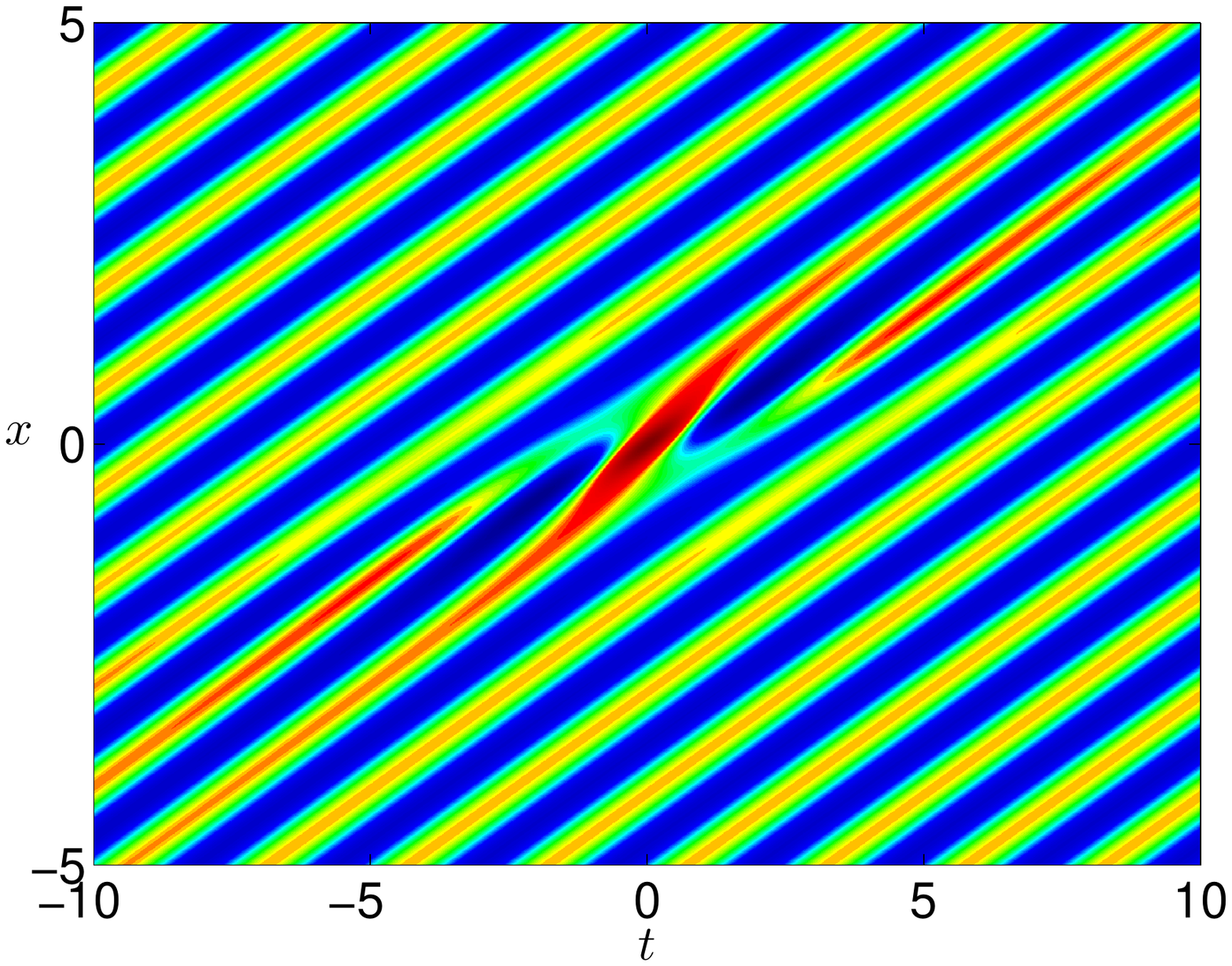}}
    \caption{\footnotesize Density plot of the physical wave field of the SFB for
    $\tilde{\nu} = 1/2$ (left), the Ma breather for $\tilde{\mu} = 0.4713$ (center)
    and the rational breather (right) corresponding to Figure \ref{3D} for $k_{0} = 2\pi$.
    The plots are shown in a moving frame of reference with suitable chosen velocity.}
    \label{physical}
  \end{center}
\end{figure}

Density plots for three breather solutions are given in Figure
\ref{physical}. For illustrations, we choose $r_{0} = 1$, $k_{0} =
2\pi$ ($\omega_{0} \approx \sqrt{2\pi}$) in all cases, modulation
frequency $\tilde{\nu} = 1/2$ for the SFB and $\tilde{\mu} =
0.4713$ for the Ma breather. For better representation, the
physical wave fields are shown in a moving frame of reference with
suitable chosen group velocity. The SFB has extreme values at $x =
\xi = 0$ and is periodic in time. The Ma breather has extreme
values at $\tau = 0$ and is periodic in space. The rational
breather is neither periodic in time nor in space, but isolated
with its maximum at $(x,t) = (0,0)$. Despite some differences, all
the breather solutions show `wavefront dislocation', when
splitting or merging of waves occurs (Nye and Berry, 1974). A
necessary condition for this phenomenon to occur is that the
Chu-Mei quotient of the nonlinear dispersion relation at the
vanishing amplitude is unbounded. When it occurs, the real-valued
phase of the corresponding wave field is undefined at the
vanishing amplitude, known as `phase singularity'. More
information on wavefront dislocation in surface water waves can be
found in (Tanaka, 1995; Karjanto and van Groesen, 2007).

\begin{center}
{\bf 6. CONCLUSIONS}
\end{center}

In this letter, we have shown that a description with displaced
phase-amplitude variables, for which the phase is time
independent, leads to the three breather solutions of the NLS
equation: the Soliton on Finite Background, the Ma breather, and
the rational breather. These are obtained as explicit solutions of
the equation for the displaced amplitude of the NLS equation. We
described the relation between these three breather solutions and
presented the amplitude amplification factor of each solution as
function of the parameters. For the corresponding physical wave
fields, wavefront dislocation and phase singularity at vanishing
amplitude are observed in all three cases.

\vspace{1.5cc} \noindent{\bf Acknowledgement.} Partly of this work
is executed at University of Twente, The Netherlands as part of
the project `Prediction and generation of deterministic extreme
waves in hydrodynamic laboratories' (TWI.5374) of the Netherlands
Organization of Scientific Research NWO, subdivision Applied
Sciences STW.

\vspace{2cc}
\begin{center}
{\small\bf REFERENCES}
\end{center}

\newcounter{ref}
\begin{list}{\small \arabic{ref}.}{\usecounter{ref} \leftmargin 4mm
\itemsep -1mm}

%Fill the list of references in the alphabetical order here. You may use
%\thebibliography  instead, and/or labels. Adopt the following style:

%for article in a journal:
\item{\small {\sc M. J. Ablowitz, et. al.}
% D. J. Kaup, A. C. Newell and H. Segur},
The inverse scattering transform-Fourier analysis for
nonlinear problems, {\it Stud. Appl. Math.}, {\bf 53}, 249-315,
1974.}

%for a book:
\item{\small {\sc M. J. Ablowitz and H. Segur}, {\it Solitons and
Inverse Scattering Transform}, Society for Industrial and Applied
Mathematics, Philadelphia, 1981.}

\item{\small {\sc M. J. Ablowitz and B. M. Herbst}, On homoclinic
structure and numerically induced chaos for the NLS equation, {\it SIAM J. Appl. Math.}, {\bf 50}, 2:339-351, 1990.}

\item{\small {\sc M. J. Ablowitz and P. A. Clarkson}, {\it
Solitons, Nonlinear Evolution Equations and Inverse Scattering},
volume \textbf{149} of {\it London Mathematical Society Lecture
Note Series}. Cambridge University Press, 1991.}

\item{\small {\sc N. N. Akhmediev, V. M.
Eleonski$\breve{\textmd{i}}$ and N. E. Kulagin}, Generation of
periodic trains of picosecond pulses in an optical fiber: exact
solutions, {\it Sov. Phys. JETP}, {\bf 62}, 5: 894-899, 1985.}

\item{\small {\sc N. N. Akhmediev and V. I. Korneev}, Modulation
instability and periodic solutions of the nonlinear
Schr\"{o}dinger equation, {\it Teoret. Mat. Fiz.}, {\bf 62},
2:189-194, 1986. English translation: {\it Theoret. Math. Phys.}
{\bf 69}, 1089-1092, 1986.}

\item{\small {\sc N. N. Akhmediev, V. M.
Eleonski$\breve{\textmd{i}}$, and N. E. Kulagin}, First-order
exact solutions of the nonlinear Schr\"{o}dinger equation, {\it
Teoret. Mat. Fiz.}, {\bf 72}, 2:183-196, 1987. English
translation: {\it Theoret. Math. Phys.}, {\bf 72}, 2:809-818,
1987.}

\item{\small {\sc N. N. Akhmediev and A. Ankiewicz}, {\it
Solitons---Nonlinear Pulses and Beams}, volume {\bf 5} of {\it
Optical and Quantum Electronic Series}, Chapman \& Hall, first
edition, 1997.}

\item{\small {\sc Andonowati, N. Karjanto and E. van Groesen},
Extreme wave phenomena in down-stream running modulated waves,
{\it Appl. Math. Model.}, {\bf 31}, 1425--1443, 2007.}

\item{\small {\sc T. B. Benjamin and J. E. Feir}, The
disintegration of wave trains on deep water. Part 1. Theory., {\it
J. Fluid Mech.}, {\bf 27}, 3:417-430, 1967.}

\item{\small {\sc A. Calini and C. M. Schober}, Homoclinic chaos
increases the likelihood of rogue wave formation, {\it Phys. Lett.
A}, {\bf 298}, 335-349, 2002.}

\item{\small {\sc P. G. Drazin and R. S. Johnson}, {\it Solitons:
an Introduction}, Cambridge University Press, 1989.}

\item{\small {\sc K. B. Dysthe and K. Trulsen}, Note on breather
type solutions of the NLS as models for freak-waves, {\it Phys.
Scripta}, {\bf T82}, 48-52, 1999}.

\item{\small {\sc K. B. Dysthe}, Modelling a ``rogue
wave"--speculations or a realistic possibility? In M. Olagnon and
G. A. Athanassoulis, editors, {\it Proceedings Rogue Waves 2000},
Ifremeer, Brest, France, 2001.}

\item{\small {\sc C. S. Gardner, J. M. Greene, M. D. Kruskal and
R. M. Miura}, Method for solving the Korteweg-de Vries equation,
{\it Phys. Rev. Lett.} \textbf{19}, 19:1095-1097, 1967.}

\item{\small {\sc R. Grimshaw, D. Pelinovsky, E. Pelinovsky and T.
Talipova}, Wave group dynamics in weakly nonlinear long-wave
models, {\it Physica D} \textbf{159}, 35-37, 2001.}

\item{\small {\sc K. L. Henderson, D. H. Peregrine, and J. W.
Dold}, Unsteady water wave modulations: fully nonlinear solutions
and comparison with the nonlinear Schr\"{o}dinger equation.
\textit{Wave Motion}, \textbf{29}, 341-361, 1999.}

\item{\small {\sc R. Hirota}, Direct method of finding exact
solutions of nonlinear evolution equations, in R. M. Miura,
editor, \textit{B\"{a}cklund Transformations, the Inverse
Scattering Method, Solitons, and Their Applications},
Springer-Verlag, Berlin, 1976.}

\item{\small {\sc R. H. M. Huijsmans, G. Klopman, N. Karjanto and
Andonowati}, Experiments on extreme wave generation using the
Soliton on Finite Background, in M. Olagnaon and M. Prevosto,
editors, {\it Proceedings Rogue Waves 2004}, Ifremer, Brest,
France, 2005.}

\item{\small {\sc N. Karjanto}, {\it Mathematical Aspects of
Extreme Water Waves}, PhD thesis, University of Twente, 2006.}

\item{\small {\sc N. Karjanto and E. van Groesen}, Note on
wavefront dislocation in surface water waves, {\it Phys. Lett. A},
{\bf 37}, 173-179, 2007.}

\item{\small {\sc Y. Li and D. W. McLaughlin}, Morse and Melnikov
functions for NLS PDE's, {\it Commun. Math. Phys.}, {\bf 162},
175-214, 1994.}

\item{\small {\sc Y. Li and D. W. McLaughlin}, Homoclinic orbits
and chaos in discretized perturbed NLS systems: Part I. Homoclinic
orbits, {\it J. Nonlinear Sci.}, {\bf 7}, 211-269, 1997.}

\item{\small {\sc Y.-C. Ma}, The perturbed plane-wave solutions of
the cubic Schr\"{o}dinger equation, {\it Stud. Appl. Math.}, {\bf
60}, 1:43-58, 1979.}

\item{\small {\sc A. Nakamura and R. Hirota}, A new example of
explode-decay solitary waves in one dimension, {\it J. Phys. Soc.
Japan}, {\bf 54}, 2:491-499, 1985.}

\item{\small {\sc J.F. Nye and M.V. Berry}, Dislocation in wave
trains, {\it Proc. R. Soc. Lond. A}, {\bf 336}, 1065:165-190,
1974.}

\item{\small {\sc M. Onorato, A. Osborne, M. Serio and T.
Damiani}, Occurrence of freak waves from envelope equations in
random ocean wave simulations, in M. Olagnon and G. A.
Athanassoulis, editors, {\it Proceedings Rogue Waves 2000},
Ifremer, Brest, France, 2001.}

\item{\small {\sc A. R. Osborne, M. Onorato and M. Serio}, The
nonlinear dynamics of rogue waves and holes in deep-water gravity
wave trains, {\it Phys. Lett. A}, {\bf 275}, 386-393, 2000.}

\item{\small {\sc D. H. Peregrine}, Water waves, nonlinear
Schr\"{o}dinger equations and their solutions, {\it J. Austral.
Math. Soc. Ser. B}, {\bf 25}, 1:16-43, 1983.}

\item{\small {\sc M. Tanaka}, Dissapearance of waves in modulated
train of surface gravity waves, {\it Structure and Dynamics of
Nonlinear Waves in Fluids}, in A. Mielke and K. Kirchg\"{a}ssner,
editors, IUTAM/ISIMM Symposium Proceedings (Hannover,
August 1994), volume {\bf 7} of {\it Advanced Series in Nonlinear
Dynamics}, World Scientific, Singapore, 392-398, 1995.}

\item{\small {\sc M. Tajiri and T. Arai}, Periodic soliton
solutions to the Davey-Stewartson equation, {\it Proc. Inst. Math.
Natl. Acad. Sci. Ukr.}, {\bf 30}, 1:210-217, 2000.}

\item{\small {E. van Groesen, Andonowati and N. Karjanto},
Displaced phase-Amplitude variables for waves on finite
background, {\it Phys. Lett. A}, {\bf 354}, 312-319, 2006.}

\item{\small {\small V. E. Zakharov and A. B. Shabat}, Exact
theory of two-dimensional self-focusing and one-dimensional
self-modulation of waves in nonlinear media, {\it Sov. Phys.
JETP}, {\bf 34}, 1:62-69, 1972.}
\end{list}

\vspace{1cc}

%Type author(s)' name, institution and e-mail address(es) here.
%For example:
{\small\noindent
{\sd Natanael Karjanto}: School of Applied Mathematics, %
The University of Nottingham Malaysia Campus, %
Jalan Broga, 43500 Semenyih, Selangor Darul Ehsan,
Malaysia.\\
\noindent E-mail: \texttt{natanael.karjanto@nottingham.edu.my}

\vspace{1cc}

\noindent{\sd E. van Groesen}: Department of Applied Mathematics, %
University of Twente, P.O. Box 217, 7500 AE Enschede, The Netherlands.\\
\noindent E-mail: \texttt{groesen@math.utwente.nl} }

\end{document}